\documentclass[preprint,showpacs,preprintnumbers,amsmath,amssymb,endfloats*]{revtex4}
\usepackage{graphicx}

\newcommand{\bes}{\begin{eqnarray}}
\newcommand{\ees}{\end{eqnarray}}

\begin{document}

\thispagestyle{empty}
\title{
Violation of the Nernst heat theorem in the theory of thermal
Casimir force between Drude metals
}
\author{V.~B.~Bezerra, 
   G.~L.~Klimchitskaya,\footnote {On leave from
North-West Technical University, St.\ Petersburg, Russia}
V.~M.~Mostepanenko,\footnote{On leave from Noncommercial
Partnership ``Scientific Instruments'', Moscow, Russia}
and C.~Romero
}

\affiliation{
Departamento de F\'{\i}sica, Universidade Federal da Para\'{\i}ba,
C.P.5008, CEP 58059--970, Jo\~{a}o Pessoa, Pb-Brazil 
}

\begin{abstract}
We give a rigorous analytical derivation of low-temperature 
behavior of the Casimir entropy in the framework of the Lifshitz
formula combined with the Drude dielectric function. An earlier
result that the Casimir entropy at zero temperature is not equal to 
zero and depends on the parameters of the system is confirmed,
i.e. the third law of thermodynamics (the Nernst heat theorem) is
violated. We illustrate the resolution of this thermodynamical 
puzzle in the context of the surface impedance approach by
several calculations of the thermal Casimir force and entropy for
both real metals and dielectrics. Different representations for
the impedances, which are equivalent for real photons, are
discussed. Finally, we argue in favor of the Leontovich
boundary condition which leads to results for the thermal
Casimir force that are consistent with thermodynamics.
\end{abstract}

\pacs{12.20.Ds, 12.20.Fv, 42.50.Lc, 05.70.-a}

\maketitle

\section{Introduction}

During the last few years the Casimir effect \cite{1} has attracted
a lot of experimental and theoretical attention as a nontrivial
macroscopic evidence for the existence of zero-point oscillations
of electromagnetic field (see, e.g., monographs \cite{2,3,4} and
reviews \cite{5,6}). Except for calculations of the Casimir force
between perfectly shaped bodies made of ideal metal at zero
temperature, the topical studies were conducted taking into account
the effects of surface roughness, finite conductivity of the
boundary metal and nonzero temperature \cite{6}.

The correct theoretical description of the thermal Casimir force
between real metals has assumed great importance due to 
recent precision experiments \cite{7,8,9,10,11,12,13,14,15,16},
followed by the prospective applications of the Casimir effect
in nanotechnology \cite{17,18} and also its use as a test for predictions
of fundamental physical theories \cite{16,19,20,21,22,23}. It was
found unexpectedly that the calculations of the thermal Casimir force
between two parallel plates made of real metal
based on the Lifshitz formula \cite{24} supplemented by
some model dielectric function ran into serious difficulties. 
The key question of the controversy is whether the transverse electric
zero mode contributes to the Casimir effect in the case of real
metals. In Refs.~\cite{25,26} a positive answer to this question was
obtained by the substitution of the plasma dielectric function into
the Lifshitz formula. In the limit of infinitely high conductivity
the results of Ref.~\cite{25,26} are smoothly transformed into the
familiar results for ideal metals \cite{6}. In Ref.~\cite{27}
the Drude dielectric function was used to calculate the contributions
of both longitudinal and transverse modes to the Casimir force.
It was found that the transverse electric zero mode does not
contribute \cite{27}. The results of Ref.~\cite{27} do not 
smoothly join with those for ideal metals. In the high-temperature
limit, where the contribution of the zero modes is dominant, the
Casimir force between plates made of Drude metals proves to be
equal to one half of the force between the ideal metals. Today,
there is an extensive literature on these two and
other approaches to the calculation of the thermal Casimir force
between real metals 
\cite{25,26,27,28,29,30,31,32,32a,33,34,35,36,37,38,38a}.

From this discussion it has been shown \cite{38} that
the substitution of the Drude dielectric function into the Lifshitz
formula results in negative values of the Casimir entropy
within wide temperature interval and leads to the violation of the
third law of thermodynamics (the Nernst heat theorem). On the
contrary, Refs.~\cite{31,32,32a} have thrown doubt on the computations
of Ref.~\cite{38} and presented the numerical computations in favor 
of the statement that the Casimir entropy 
for the Drude metals is zero at
zero temperature, i.e. the third law of thermodynamics is not
violated.

In the present paper we give a detailed and rigorous analytical 
derivation of the low-temperature behavior of the Casimir entropy in 
the framework of the Lifshitz formula combined with the Drude dielectric
function (Sec.~II). This derivation removes the doubts raised in 
Refs.~\cite{31,32,32a} and validates the thermodynamic inconsistency of
the Drude dielectric function with the Lifshitz formula at nonzero
temperature. The reason why the opposite conclusion is obtained in
Refs.~\cite{31,32,32a} is explained in Sec.~III. This section contains
also the resolution of the above thermodynamic puzzle 
by presenting
several computations in the framework of the surface impedance
approach \cite{35,36} as opposed to the use of the Drude model.
The exact boundary conditions in terms of impedances, depending
on polarization and angle of incidence, are compared with the
Leontovich boundary condition. The latter is shown to be applicable
to the case of fluctuating fields being in agreement with the third
law of thermodynamics.
The temperature dependences of the Casimir force and entropy for
real metals in both approaches are compared with those for dielectrics. 
Sec.~IV contains conclusions and discussion of
recent experimental results.

\section{Casimir free energy and entropy in the Lifshitz theory
combined with the Drude model}

The Casimir free energy for the configuration of two parallel plates at
a separation $a$ and temperature $T$ is given by the Lifshitz
formula \cite{24}, which can be represented in the form
\begin{equation}
{\cal{F}}(a,T)=\frac{k_BT}{8\pi a^2}
\sum\limits_{l=0}^{\infty}{\vphantom{\sum}}^{\prime}
\int_{\zeta_l}^{\infty}ydy
\left\{\ln\left[1-r_{\|}^2(\zeta_l,y)e^{-y}\right]
+\ln\left[1-r_{\bot}^2(\zeta_l,y)e^{-y}\right]\right\}.
\label{eq1}
\end{equation}
\noindent
Here {\it prime} adds a multiple 1/2 near the term with $l=0$, $k_B$ is
the Boltzmann constant, the dimensionless Matsubara frequencies
are $\zeta_l=\xi_l/\omega_c$, where $\xi_l=2\pi k_B Tl/\hbar$,
$\omega_c=c/(2a)$, and the reflection coefficients for the two
different polarizations are expressed in terms of the dielectric
permittivity $\varepsilon(\omega)$ as
\begin{eqnarray}
&&
r_{\|}^2(\zeta_l,y)=
\left\{\frac{y\varepsilon(i\zeta_l)-
\sqrt{\left[\varepsilon(i\zeta_l)-1\right]\zeta_l^2+
y^2}}{y\varepsilon(i\zeta_l)+
\sqrt{\left[\varepsilon(i\zeta_l)-1\right]\zeta_l^2+
y^2}}\right\}^2,
\nonumber \\
&&
r_{\bot}^2(\zeta_l,y)=
\left\{\frac{y-
\sqrt{\left[\varepsilon(i\zeta_l)-1\right]\zeta_l^2+
y^2}}{y+
\sqrt{\left[\varepsilon(i\zeta_l)-1\right]\zeta_l^2+
y^2}}\right\}^2.
\label{eq2}
\end{eqnarray}

We consider the Drude metals described by the dielectric permittivity
of the Drude model. At the imaginary Matsubara frequencies it is
given by
\begin{equation}
\varepsilon^{(D)}(i\zeta_l)=1+
\frac{{\tilde{\omega}}_p^2}{\zeta_l\left[\zeta_l+{\tilde{\gamma}}(T)\right]}, 
\label{eq3}
\end{equation}
\noindent
where ${\tilde{\omega}}_p$ and ${\tilde{\gamma}}(T)$ are the dimensionless
plasma frequency and relaxation parameter  defined by
${\tilde{\omega}}_p=\omega_p/\omega_c$, 
${\tilde{\gamma}}(T)=\gamma(T)/\omega_c$.
In the absence of relaxation ${\tilde{\gamma}}(T)=0$ and 
$\varepsilon^{(D)}$ coincides with the dielectric permittivity
of the free electron plasma model $\varepsilon^{(p)}$.
Substituting Eq.~(\ref{eq3}) into Eqs.~(\ref{eq1}), (\ref{eq2}) one
obtains the Casimir free energy ${\cal{F}}^{(D)}(a,T)$ and the reflection
coefficients $r_{\|,\bot}^{(D)}(\zeta_l,y)$ in the framework of
the Drude model. If, from the very beginning, $\gamma(T)=0$, then
the Casimir free energy in the framework of the plasma model
${\cal{F}}^{(p)}(a,T)$ and reflection
coefficients $r_{\|,\bot}^{(p)}(\zeta_l,y)$ are obtained.
It is notable that $r_{\bot}^{(D)}(0,y)=0$, whereas
\begin{equation}
{r_{\bot}^{(p)}}^2(0,y)=
{r_{\bot}^{(p)}}^2(\zeta_l,y)=
\left(\frac{y-
\sqrt{{\tilde{\omega}}_p^2+y^2}}{y+
\sqrt{{\tilde{\omega}}_p^2+y^2}}\right)^2\neq 0,
\label{eq3a}
\end{equation}
i.e. there is no smooth transition from ${\cal{F}}^{(D)}$ to
${\cal{F}}^{(p)}$ when $\gamma(T)\to 0$.
This nonanalyticity is determined exclusively by the zero-frequency
contribution of the transverse electric mode to the free energy
${\cal{F}}^{(D)}(a,T)$.

For the calculation of the Drude free energy ${\cal{F}}^{(D)}(a,T)$
it is useful to represent it as the plasma free energy
${\cal{F}}^{(p)}(a,T)$ plus some additional terms.
Taking into account that $r_{\bot}^{(D)}(0,y)=0$ and
${r_{\|}^{(p)}}^2(0,y)={r_{\|}^{(D)}}^2(0,y)=1$, this identical
representation is as follows
\begin{eqnarray}
&&
{\cal{F}}^{(D)}(a,T)={\cal{F}}^{(p)}(a,T)-\frac{k_BT}{16\pi a^2}
\int_{0}^{\infty}ydy\ln\left[1-{r_{\bot}^{(p)}}^2(0,y)e^{-y}\right]
\label{eq4} \\
&&\phantom{aaa}+
\frac{k_BT}{8\pi a^2}
\sum\limits_{l=1}^{\infty}{\vphantom{\sum}}^{\prime}
\int_{\zeta_l}^{\infty}ydy
\left\{\ln\left[1-{r_{\|}^{(D)}}^2(\zeta_l,y)e^{-y}\right]
-\ln\left[1-{r_{\|}^{(p)}}^2(\zeta_l,y)e^{-y}\right]\right.
\nonumber \\
&&\phantom{aaaaaa}\left.
+\ln\left[1-{r_{\bot}^{(D)}}^2(\zeta_l,y)e^{-y}\right]
-\ln\left[1-{r_{\bot}^{(p)}}^2(\zeta_l,y)e^{-y}\right]\right\}.
\nonumber
\end{eqnarray}
\noindent
An important point is that the zero-frequency contributions
are contained only in the first two terms of the right-hand side
of Eq.~(\ref{eq4}), whilst the summation starts from the term
with $l=1$ leading to nonzero lower integration limits $\zeta_l$
in the integrals at any nonzero temperature.

It is instructive to find the asymptotic representation for
the free energy (\ref{eq4}) applicable at $T\to 0$.
First we notice that among the three parameters, contained
in Eq.~(\ref{eq3}), i.e ${\tilde{\omega}}_p$, $\zeta_l$
(with $l\geq 1$) and ${\tilde{\gamma}}(T)$, the latter is the 
smallest one. In fact at $T=300\,$K for good metals
$\gamma\sim 10^{13}-10^{14}\,$rad/s (for Au 
$\gamma=5.32\times 10^{13}\,$rad/s) whereas
$\xi_1=2\pi k_BT/\hbar=2.46\times 10^{14}\,$rad/s, and
$\xi_l=l\xi_1$, i.e. in all cases $\gamma<\xi_l$.
When $T$ decreases from room temperature up to approximately $T_D/4$, 
where $T_D$ 
is the Debye temperature ($T_D=165\,$K for Au \cite{39}),
$\gamma(T)\sim T$, i.e. $\gamma(T)$ decreases following the same law 
as $\xi_l$, preserving the inequality $\gamma(T)<\xi_l$.
At $T<T_D/4$ the relaxation parameter decreases even more
quickly than $\xi_l$ with decreasing $T$ (as $\sim T^5$
according to the Bloch-Gr\"{u}neisen 
law due to electron-phonon collisions
\cite{40} and as $\sim T^2$ at liquid helium temperatures
due to electron-electron scattering \cite{39}).
As a result, 
at $T=30\,$K
$\gamma(T)/\xi_1(T)\approx 4.9\times 10^{-2}$,
at $T=10\,$K $\gamma(T)/\xi_1(T)\approx 1.8\times 10^{-3}$,
and this relation decreases further with $T\to 0$, i.e. at low 
temperatures the condition $\gamma(T)\ll\xi_l(T)$ is largely 
satisfied.

The largest parameter of the above three is $\omega_p$
(for Au $\omega_p=1.37\times 10^{16}\,$rad/s).
For example, for Au at $T=300\,$K, 70\,K, and 10\,K we have 
$\gamma(T)/\omega_p={\tilde{\gamma}}(T)/{\tilde{\omega}}_p=
3.88\times 10^{-3}$,
$6.71\times 10^{-4}$, $1.06\times 10^{-6}$, respectively.

The reflection coefficients (\ref{eq2}) have continuous derivatives
with respect to the relation $\gamma(T)/\xi_l(T)$ ($l\geq 1$) at 
the point $\gamma(T)/\xi_l(T)=0$.
Under the above proved condition $\gamma(T)\ll\xi_l(T)$, which is
satisfied at all sufficiently low temperatures, the relation
$\gamma(T)/\xi_l(T)\ll 1$, and we can
expand in Taylor series around a point $\gamma(T)/\xi_l(T)=0$ 
keeping only the first order terms
\begin{eqnarray}
&&
{r_{\|}^{(D)}}^2(\zeta_l,y)=
{r_{\|}^{(p)}}^2(\zeta_l,y)-\frac{{\tilde{\gamma}}(T)}{\zeta_l(T)}
R_{\|}(\zeta_l,y),
\nonumber \\
&&
{r_{\bot}^{(D)}}^2(\zeta_l,y)=
{r_{\bot}^{(p)}}^2(\zeta_l,y)-\frac{{\tilde{\gamma}}(T)}{\zeta_l(T)}
R_{\bot}(\zeta_l,y),
\label{eq5}
\end{eqnarray}
\noindent
where
\begin{eqnarray}
&&
R_{\|}(\zeta_l,y)=\frac{2\zeta_l^2\alpha y\left[1+\alpha^2\left(2y^2-
\zeta_l^2\right)\right]|r_{\|}^{(p)}(\zeta_l,y)
|}{\sqrt{1+\alpha^2y^2}\left[y+\alpha\zeta_l^2\left(\alpha y+
\sqrt{1+\alpha^2y^2}\right)\right]^2},
\nonumber \\
&&
R_{\bot}(\zeta_l,y)=
\frac{2\alpha y|r_{\bot}^{(p)}(\zeta_l,y)|}{\sqrt{1+
\alpha^2y^2}\left(\alpha y+
\sqrt{1+\alpha^2y^2}\right)^2},
\label{eq6}
\end{eqnarray}
\noindent
and $\alpha\equiv 1/{\tilde{\omega}}_p=\lambda_p/(4\pi a)$,
$\lambda_p$ is the plasma wavelength (note that we keep the
argument $T$ when $\zeta_l$ participates in the expansion
parameter but omit it in the other cases).

The same expansions for the logarithms which appear in Eq.~(\ref{eq1}) are
\begin{eqnarray}
&&
\ln\left[1-{r_{\|}^{(D)}}^2(\zeta_l,y)e^{-y}\right]=
\ln\left[1-{r_{\|}^{(p)}}^2(\zeta_l,y)e^{-y}\right]
+\frac{{\tilde{\gamma}}(T)}{\zeta_l(T)}
\frac{R_{\|}(\zeta_l,y)e^{-y}}{1-{r_{\|}^{(p)}}^2(\zeta_l,y)e^{-y}},
\nonumber \\
&&
\label{eq7} \\
&&
\ln\left[1-{r_{\bot}^{(D)}}^2(\zeta_l,y)e^{-y}\right]=
\ln\left[1-{r_{\bot}^{(p)}}^2(\zeta_l,y)e^{-y}\right]
+\frac{{\tilde{\gamma}}(T)}{\zeta_l(T)}
\frac{R_{\bot}(\zeta_l,y)e^{-y}}{1-{r_{\bot}^{(p)}}^2(\zeta_l,y)e^{-y}}.
\nonumber
\end{eqnarray}

Substituting Eq.~(\ref{eq7}) in Eq.~(\ref{eq4}) we obtain the following
expression 
for the Casimir free energy between parallel plates made of Drude metal
\begin{equation}
{\cal{F}}^{(D)}(a,T)={\cal{F}}^{(p)}(a,T)
-\frac{k_BT}{16\pi a^2}
\int_{0}^{\infty}ydy\ln\left[1-{r_{\bot}^{(p)}}^2(0,y)e^{-y}\right]
+{\cal{F}}^{(\gamma)}(a,T),
\label{eq8}
\end{equation}
\noindent
where the contribution depending on the relaxation parameter is 
given by
\begin{equation}
{\cal{F}}^{(\gamma)}(a,T)=
\frac{{\tilde{\gamma}}(T)}{{\tilde{\omega}}_p}\frac{k_BT}{8\pi a^2}
\sum\limits_{l=1}^{\infty}
\frac{{\tilde{\omega}}_p}{\zeta_l(T)}
\int_{\zeta_l}^{\infty}ydy\left[
\frac{R_{\|}(\zeta_l,y)}{e^{y}-{r_{\|}^{(p)}}^2(\zeta_l,y)}
+\frac{R_{\bot}(\zeta_l,y)}{e^{y}-{r_{\bot}^{(p)}}^2(\zeta_l,y)}
\right].
\label{eq9}
\end{equation}
\noindent
Notice that in this expression the small parameter
${\tilde{\gamma}}(T)/{\tilde{\omega}}_p=\gamma(T)/\omega_p$,
which does not depend on an index of summation, is put in evidence.

The low-temperature asymptotic limit of the Casimir free energy 
${\cal{F}}^{(p)}(a,T)$ in the framework of the plasma model
was investigated with details in Ref.~\cite{38} (the coinciding
numerical results follow also from the computations of
Ref.~\cite{25}). In terms of the two small parameters $T/T_{eff}$
[where $k_BT_{eff}\equiv\hbar c/(2a)=\hbar\omega_c$] and
$\delta_0/a=2\alpha$ [where $\delta_0=\lambda_p/(2\pi)$ is the
skin layer thickness in the frequency region of the infrared
optics] the plasma model free energy is given by \cite{38}
\begin{equation}
{\cal{F}}^{(p)}(a,T)=E^{(p)}(a)-\frac{\hbar c\zeta(3)}{16\pi a^3}
\left[
\vphantom{\left[\left(\frac{T}{T_{eff}}\right)^2e^{2\pi T_{eff}/T}\right]}
\left(1+2\frac{\delta_0}{a}\right)\left(\frac{T}{T_{eff}}\right)^3
-\frac{\pi^3}{45\zeta(3)}
\left(1+4\frac{\delta_0}{a}\right)\left(\frac{T}{T_{eff}}\right)^4
\right],
\label{eq10}
\end{equation}
\noindent
where $E^{(p)}(a)$ is the Casimir energy at zero temperature
calculated by using the plasma model dielectric function.
The perturbation expansion (\ref{eq10}) is applicable at separations
$\lambda_p\leq a<3\,\mu$m at all temperatures $T\leq 300\,$K \cite{26}.

The second term in the right-hand side of Eq.~(\ref{eq8}) is linear
in the temperature. It is an easy matter to calculate the coefficient
near $T$ perturbatively. For this purpose we use Eq.(\ref{eq3a}) and
expand the logarithm under the integral in powers of $\delta_0/a$. Then
all integrals are taken explicitly, resulting in
\begin{eqnarray}
&&
-\frac{k_BT}{16\pi a^2}
\int_{0}^{\infty}ydy\ln\left[1-{r_{\bot}^{(p)}}^2(0,y)e^{-y}\right]=
\frac{k_BT\zeta(3)}{16\pi a^2}
\left\{
\vphantom{O\left[\left(\frac{\delta_0}{a}\right)^5\right]}
1-4\frac{\delta_0}{a}+
12\left(\frac{\delta_0}{a}\right)^2\right.
\label{eq11} \\
&&
\phantom{aaaa}
\left.
-32\left(\frac{\delta_0}{a}\right)^3
\left[1-\frac{\zeta(5)}{16\zeta(3)}\right]
+80\left(\frac{\delta_0}{a}\right)^4
\left[1-\frac{\zeta(5)}{4\zeta(3)}\right]\right\},
\nonumber
\end{eqnarray}
\noindent
where $\zeta(z)$ is the Riemann zeta function.

Now we consider the low-temperature behavior of the last term
in the right-hand side of Eq.~(\ref{eq8}),
${\cal{F}}^{(\gamma)}(a,T)$, which depends on the relaxation
parameter. An important point is that the relaxation parameter
is not an independent one but is the function of the temperature,
$\gamma=\gamma(T)$. Because of this, it is not improbable that
the temperature dependent terms resulting from the integration
and summation in Eq.~(\ref{eq9}) will cancel the second term, 
linear in the temperature, in the Casimir free energy (\ref{eq8}).
Below we demonstrate that this is not the case.

As pointed out above, $\gamma(T)\to 0$ when $T\to 0$ no slower 
than $\sim T^2$. If desired that the quantity 
${\cal{F}}^{(\gamma)}(a,T)$ from Eq.~(\ref{eq9}) be linear in $T$,
the sum in Eq.~(\ref{eq9}) should tend to infinity as
$1/T^2$ when $T\to 0$. Let us find what is the actual asymptotic
behavior of the quantity 
${\cal{F}}^{(\gamma)}(a,T)$ when $T\to 0$. 
For this purpose we expand $R_{\|,\bot}(\zeta_l,y)$ from Eq.~(\ref{eq6})
up to the first order in the small parameter $\alpha$ (recall that
${\cal{F}}^{(\gamma)}$ is already proportional to the smallest parameter
of our problem $\gamma/\omega_p$)
\begin{equation}
R_{\|}(\zeta_l,y)=\frac{2\zeta_l^2\alpha}{y},
\qquad
R_{\bot}(\zeta_l,y)=2y\alpha.
\label{eq12}
\end{equation}
\noindent
Substituting Eq.~(\ref{eq12}) into Eq.~(\ref{eq9})
one obtains
\begin{equation}
{\cal{F}}^{(\gamma)}(a,T)=\frac{\gamma(T)}{\omega_p}
\frac{k_BT}{4\pi a^2}
\sum\limits_{l=1}^{\infty}\left(
\zeta_l\int_{\zeta_l}^{\infty}\frac{dy}{e^{y}-1}+
\frac{1}{\zeta_l}
\int_{\zeta_l}^{\infty}\frac{y^2dy}{e^{y}-1}\right).
\label{eq13}
\end{equation}
\noindent
Each of the two sums in Eq.~(\ref{eq13}) can be simply
found when $T\to 0$. The asymptotic of the first sum is as follows
\begin{equation}
\sum\limits_{l=1}^{\infty}
\zeta_l\int_{\zeta_l}^{\infty}\frac{dy}{e^{y}-1}=
\frac{2\pi T}{T_{eff}}\sum\limits_{k=1}^{\infty}
\frac{1}{k}\left[\frac{1}{e^{2\pi kT/T_{eff}}-1}+
\frac{1}{\left(e^{2\pi kT/T_{eff}}-1\right)^2}\right]
\approx \frac{T_{eff}}{2\pi T}\zeta(3)+\zeta(2)
\label{eq14}
\end{equation}
\noindent
(here we have neglected all terms in the expansion of
$\exp(2\pi kT/T_{eff})$ in the denominators starting from
the third ones).
{}For the second sum in Eq.(\ref{eq13}) one obtains
\begin{eqnarray}
&&
\sum\limits_{l=1}^{\infty}
\frac{1}{\zeta_l}
\int_{\zeta_l}^{\infty}\frac{y^2dy}{e^{y}-1}=
\frac{T_{eff}}{2\pi T}\left[
\vphantom{\frac{e^{2\pi kT/T_{eff}}}{\left(e^{2\pi kT/T_{eff}}-1\right)^2}}
-2\sum\limits_{k=1}^{\infty}\frac{1}{k^3}
\ln\left(1-e^{-2\pi kT/T_{eff}}\right)+
\frac{4\pi T}{T_{eff}}\sum\limits_{k=1}^{\infty}\frac{1}{k^2}
\frac{1}{e^{2\pi kT/T_{eff}}-1}\right.
\nonumber \\
&&
\label{eq15} \\
&&
\phantom{aaa}\left.
+\frac{4\pi^2 T^2}{T_{eff}^2}\sum\limits_{k=1}^{\infty}\frac{1}{k}
\frac{e^{2\pi kT/T_{eff}}}{\left(e^{2\pi kT/T_{eff}}-1\right)^2}\right]
\approx -\frac{T_{eff}\zeta(3)}{\pi T}\ln\frac{2\pi T}{T_{eff}}+
\frac{3T_{eff}\zeta(3)}{2\pi T}+2\zeta(2).
\nonumber
\end{eqnarray}

Substituting Eqs.~(\ref{eq14}), (\ref{eq15}) into Eq.~(\ref{eq13})
we arrive at
\begin{equation}
{\cal{F}}^{(\gamma)}(a,T)\approx\frac{\gamma(T)}{\omega_p}
\frac{k_BT_{eff}\zeta(3)}{4\pi^2 a^2}\left[
-\ln\frac{2\pi T}{T_{eff}}+2+3\pi\frac{\zeta(2)}{\zeta(3)}
\frac{T}{T_{eff}}\right].
\label{eq16}
\end{equation}
\noindent
As is seen from Eq.~(\ref{eq16}), the leading term in 
${\cal{F}}^{(\gamma)}(a,T)$ behaves as $-\gamma(T)\ln(T/T_{eff})$
and goes to zero when $T\to 0$ because $\gamma(T)\sim T^2$ at
helium and lower temperatures.

Now we are in a position to find the low temperature behavior
of the Casimir entropy
\begin{equation}
S^{(D)}(a,T)=-\frac{\partial{\cal{F}}^{(D)}(a,T)}{\partial T}
\label{eq17}
\end{equation}
\noindent
calculated by using the Drude dielectric function, where the
Casimir free energy ${\cal{F}}^{(D)}(a,T)$ is given by Eq.~(\ref{eq8}).
According to Eq.~(\ref{eq8}), there are three contributions into
the Casimir entropy in the framework of the Drude model at low
temperatures. The first one is given by the Casimir entropy 
calculated by means of the free electron plasma dielectric
function. It is obtained from Eq.~(\ref{eq10})
\begin{eqnarray}
&&
S^{(p)}(a,T)=-\frac{\partial{\cal{F}}^{(p)}(a,T)}{\partial T}
=\frac{3k_B\zeta(3)}{8\pi a^2}\left(\frac{T}{T_{eff}}\right)^2
\label{eq18} \\
&&
\phantom{aaa}\times
\left\{1-\frac{4\pi^3}{135\zeta(3)}\frac{T}{T_{eff}}+
2\frac{\delta_0}{a}\left[1-
\frac{8\pi^3}{135\zeta(3)}\frac{T}{T_{eff}}\right]\right\}.
\nonumber
\end{eqnarray}
\noindent
This result coincides with the one obtained in Ref.~\cite{38}
(it is also in agreement with the numerical computations of the
thermal corrections to the Casimir force in Refs.~\cite{25,26}).
Evidently the plasma model Casimir entropy is positive and
$S^{(p)}(a,T)\to 0$ when $T\to 0$, i.e. it is in agreement with 
the Nernst heat theorem.

The second contribution to the Drude model Casimir entropy is
obtained from the second term in the right-hand side of Eq.~(\ref{eq8}) 
taking into account Eq.~(\ref{eq11}) and is given by
\begin{eqnarray}
&&
S_0^{(D)}(a,T)=S_0^{(D)}(a,0)=\frac{k_B}{16\pi a^2}
\int_{0}^{\infty}ydy\ln\left[1-{r_{\bot}^{(p)}}^2(0,y)e^{-y}\right]
\nonumber \\
&&
\phantom{aaaa}
=-\frac{k_B\zeta(3)}{16\pi a^2}\left[1-4\frac{\delta_0}{a}+12
\left(\frac{\delta_0}{a}\right)^2-\ldots\right]
\label{eq19}
\end{eqnarray}
\noindent
[see Eq.~(\ref{eq11}) for higher perturbation orders in
$\delta_0/a$].
This contribution is negative and does not depend on the temperature.

The asymptotic behavior of the last contribution to the Casimir entropy 
in the Drude model, as given by Eq.~(\ref{eq8}), at $T\to 0$, is obtained
from Eq.~(\ref{eq16}) with account $\gamma(T)=\gamma_0T^2$
\begin{equation}
S^{(\gamma)}(a,T)=-\frac{\partial{\cal{F}}^{(\gamma)}(a,T)}{\partial T}
\approx -\frac{k_B\zeta(3)}{4\pi^2 a^2}\frac{\gamma(T)}{\omega_p}
\frac{T_{eff}}{T}
\left[-2\ln\frac{2\pi T}{T_{eff}}+3+9\pi\frac{\zeta(2)}{\zeta(3)}
\frac{T}{T_{eff}}\right].
\label{eq20}
\end{equation}
\noindent
{}From Eq.~(\ref{eq20}) we notice that $S^{(\gamma)}(a,T)\to 0$
when $T\to 0$.

As a result, the value of the Casimir entropy at zero temperature
calculated with the help of the Drude dielectric function is
found from Eqs.~(\ref{eq18})--(\ref{eq20}):
\begin{equation}
S^{(D)}(a,0)=\lim\limits_{T\to 0}\left[
S^{(p)}(a,T)+S_0^{(D)}(a,T)+S^{(\gamma)}(a,T)\right]=
S_0^{(D)}(a,0)<0,
\label{eq21}
\end{equation}
\noindent
where the quantity in the right-hand side is negative and is given
by Eq.~(\ref{eq19}).
This quantity depends on the parameters of the system, such as 
the separation between the plates and plasma frequency, violating
the third law of thermodynamics \cite{41,42} (the Nernst
heat theorem). Therefore we may conclude that the Drude
dielectric function is thermodynamically inconsistent with the
Lifshitz formula and unusable to calculate the thermal
Casimir force between real metals. Note also that according to
Eqs.~(\ref{eq19})--(\ref{eq21}) the Casimir entropy between 
two parallel plates made of Drude
metals is negative within a wide separation range.
The entropy cannot be made positive by introducing some composite
system containing a subsystem with negative entropy between the two 
infinite plates without changing the value
of the Casimir force.

In the next section we discuss the possibilities to avoid the
above thermodynamical puzzle and formulate the approach
in which the Casimir entropy is positive, becoming zero 
at zero temperature.

\section{Towards a thermodynamically consistent theory of the
thermal Casimir force between real metals}

The above asymptotic limit for the Casimir entropy was derived under
the important condition that $\gamma(T)\ll\xi_l(T)$, i.e. at
any $T$ the magnitude of the relaxation parameter must be much less
than the Matsubara frequencies. If this condition does not hold,
the obtained conclusion concerning the thermodynamical
inconsistency of the Drude model combined with the Lifshitz formula
is open to question. This explains why in Refs.~\cite{31,32,32a}
it was concluded that the Lifshitz formula combined with the
Drude model respects the Nernst heat theorem. In Refs.~\cite{31,32,32a}
all computations were performed under the condition that the relaxation
parameter is constant (the values $\gamma=35.6\,$meV
or 0.01\,meV, as at $T=300\,$K and $T=10\,$K, respectively,
were extended to all temperatures).
Then even for the smaller value, at temperatures $T<0.018\,$K
the inequality $\gamma >\xi_1(T)$ is fulfilled, i.e. the condition
$\gamma(T)\ll\xi_l(T)$ is violated, and our proof in Sec.~II is
not applicable. 

The question arises whether there are physical prerequisites
for violating the condition $\gamma(T)\ll\xi_l(T)$. 
According to Sec.~II, the
nonelastic processes of electron-phonon collisions and also
the elastic electron-electron scattering respect the inequality
$\gamma(T)\ll\xi_l(T)$. One may hope, however, that some fine
properties of real metal bodies could result in the violation
of this inequality at some sufficiently low temperatures.
It has been proposed \cite{43} that this role can be played by
impurities and defects which lead to a nonzero residual value
of the static resistivity (and, thus, a relaxation parameter; recall
that resistivity is proportional to the relaxation parameter \cite{40}) 
as the temperature goes to zero (see also Appendix D in Ref.~\cite{32}).

Here we adduce the argument that impurities
cannot remedy the situation with the violation of the
Nernst heat theorem. In fact, the resistivity ratio of a sample
can be defined as the ratio of its resistivity at room temperature
to its residual resistivity. For pure samples the
resistivity ratio may be as high as $10^6$ \cite{39}.
As an example, let us consider Au with 
$\gamma(T=300\,\mbox{K})=5.32\times 10^{13}\,$rad/s.
In this case for the residual value of the relaxation parameter
one obtains $\gamma_{res}=5.32\times 10^7\,$rad/s.
The asymptotic expressions of Sec.~II are applicable under the
condition $\gamma\ll\xi_1=2\pi k_BT/\hbar$.
Thus, with allowance made for impurities, these 
asymptotics are applicable at temperatures 
$T\gg\hbar\gamma_{res}/(2\pi k_B)=6.5\times10^{-5}\,$K. 
What this means is that the Casimir entropy 
at temperature $T=5\times 10^{-4}\,$K has a nonzero negative
value given by Eq.~(\ref{eq19}). Physically this is equivalent to
the violation of the Nernst heat theorem. 
Only at smaller temperatures of about $10^{-4}\,$K does Eq.~(\ref{eq19})
break down and the Casimir entropy rapidly takes a sharp upward
turn to zero. We would like to
point out also that the usual theory of the electron-phonon
interaction, describing electrons interacting with
elementary excitations of a perfect lattice with no impurities,
must satisfy and does satisfy all the requirements of
thermodynamics. In fact, the mean free path of the electrons
between the collisions with impurities at zero temperature is many 
orders of magnitude greater than the penetration depth of the electromagnetic
oscillations at the characteristic frequency into the metal.
To make sure that this is the case, recall that the relaxation
time for Au (the inverse of the relaxation parameter) at $T=300\,$K
is equal to $\tau(T=300\,\mbox{K})=1.88\times 10^{-14}\,$s.
Using the above resistivity ratio, one obtains the value of
the relaxation time at zero temperature
$\tau(T\approx 0\,\mbox{K})=1.88\times 10^{-8}\,$s.
Finally, using the value of Fermi velocity for Au 
($v_F=1.79\times 10^{6}\,$m/s) the mean free path of the
electron is equal to 
$l(T\approx 0\,\mbox{K})=v_F\tau(T\approx 0\,\mbox{K})=3.36\,$cm.
This should be compared with the thickness of the skin layer in
the region of the infrared optics, equal to approximately 22\,nm.
That is why the attempt to remedy the violation
of the Nernst heat theorem at the expense of impurities is
meaningless.

Recently the resolution of these complicated problems was obtained
\cite{36} using another approach to the description of
real metals based on the concept of the Leontovich impedance
boundary conditions.
This approach offers a fundamental understanding of the reason why
the Drude model is not compatible with the theory of the thermal
Casimir force between real metals.

The main concept of the Lifshitz theory is the fluctuating electromagnetic
field considered on the background of dielectric permittivity
depending only on frequency. This concept works good in the case
of dielectrics but is not adequate for real metals. In fact, in the
frequency region of the anomalous skin effect the spatial
non-uniformity of the field makes impossible a description of a metal
in terms of $\varepsilon(\omega)$ \cite{44}. Then the electromagnetic
fluctuations also cannot be considered on this background.
Moreover, in the frequency region of the normal skin effect the electric
field {\boldmath$E$} initiates a real current of conduction
electrons {\boldmath$j$}$=\sigma_0${\boldmath$E$}, where $\sigma_0$ is
the DC conductivity of a metal. These
{\boldmath$j$} and {\boldmath$E$} should be considered as real ones
\cite{36}.
In contrast with {\boldmath$E$}, the fluctuating field
cannot heat a metal as {\boldmath$E$} does due to collisions of
conduction electrons with phonons.

At present a complete theory of field quantization inside metals
which, among other things, should take into account the effects of
spatial non-uniformity, is not available. Because of this, the concept
of the electromagnetic fluctuations inside a metal remains unclear.
In the absence of a complete theory we should not take into
consideration the metal interior, but rather take into account the
realistic material properties by means of the surface impedance function.

It is well known that for a plane wave of a single frequency
inside a medium with dielectric permittivity $\varepsilon$ the
following equations are valid \cite{44}
\begin{equation}
\omega{\mbox{\boldmath$H$}}=c{\mbox{\boldmath$k$}}
\times{\mbox{\boldmath$E$}}, \qquad
\omega\varepsilon{\mbox{\boldmath$E$}}=-c{\mbox{\boldmath$k$}}
\times{\mbox{\boldmath$H$}},
\label{eq22}
\end{equation}
\noindent
where {\boldmath$k$} is a complex wave vector. Then from the first
equality of Eq.~(\ref{eq22}) for the field
with transverse polarization ({\boldmath$E$} is perpendicular
to the $xz$-plane which is the plane of incidence) inside a metal
near its boundary plane  it follows
\begin{equation}
{\mbox{\boldmath$E$}}_t=Z_{\bot}\left[{\mbox{\boldmath$H$}}_t
\times{\mbox{\boldmath$n$}}\right], \qquad
Z_{\bot}=\frac{\omega}{\sqrt{\omega^2\varepsilon-c^2k_{\bot}^2}}.
\label{eq23}
\end{equation}
\noindent
Here the index $t$ refers to the component of the field parallel to 
the boundary plane
and the unit vector {\boldmath$n$} is perpendicular to it and
directed inside a metal. In the same way, from the second equality
of Eq.~(\ref{eq22}) for the field with longitudinal
polarization  one obtains
\begin{equation}
\left[{\mbox{\boldmath$n$}}
\times{\mbox{\boldmath$E$}}_t\right]=Z_{\|}
{\mbox{\boldmath$H$}}_t, \qquad
Z_{\|}=\frac{1}{\omega\varepsilon}\sqrt{\omega^2\varepsilon-c^2k_{\bot}^2}.
\label{eq24}
\end{equation}
\noindent
The quantities $Z_{\bot,\|}$ are called impedances \cite{45}. 
By the use of the Snell's law
they can be identically represented as
\begin{equation}
Z_{\bot}=\frac{1}{\sqrt{\varepsilon}\sqrt{1-
\frac{\sin^2\vartheta_0}{\varepsilon}}}, \qquad
Z_{\|}=\frac{1}{\sqrt{\varepsilon}}\sqrt{1-
\frac{\sin^2\vartheta_0}{\varepsilon}},
\label{eq25}
\end{equation}
\noindent
where $\vartheta_0$ is the angle of incidence of the electromagnetic wave
from vacuum on the boundary plane of the metal.

In metals for all frequencies which are at least several times less
than the plasma frequency we have that $|\varepsilon|\gg 1$. For this 
reason, the term $\sin^2\vartheta_0/\varepsilon$ in Eqs.~(\ref{eq25}) can be 
neglected in comparison with unity. This leads to the fact that inside a 
metal all waves are spreaded perpendicular to the surface, i.e. the
refraction angle is equal to zero independently of the angle of
incidence \cite{44}. As Leontovich has suggested \cite{44}, the equations
\begin{equation}
{\mbox{\boldmath$E$}}_t=Z\left[{\mbox{\boldmath$H$}}_t
\times{\mbox{\boldmath$n$}}\right], \qquad
\left[{\mbox{\boldmath$n$}}
\times{\mbox{\boldmath$E$}}_t\right]=Z
{\mbox{\boldmath$H$}}_t
\label{eq26}
\end{equation}
\noindent
with $Z=1/\sqrt{\varepsilon}$ can be used as boundary conditions in order
to determine the field outside the metal. The quantity $Z=Z(\omega)$
is called the impedance of a metal \cite{44} or the ``intrinsic''
impedance \cite{45}. It depends only on $\omega$ and does not depend on
the polarization or the angle of incidence. 
We emphasize that for real photons the difference between the Leontovich
impedance, as is in Eq.~(\ref{eq26}), and impedances in Eq.~(\ref{eq25})
is negligibly small. What is more, when $\omega\to 0$, the dielectric
permittivity goes to infinity and the Leontovich impedance coincides
precisely with the impedances (\ref{eq25}). By postulating the boundary
condition (\ref{eq26}) in the theory of the Casimir effect, we admit,
in fact, that the virtual photons have the same reflection properties
on the metal boundary as real ones do.
It is significant that the surface impedance and 
the boundary conditions (\ref{eq26}) still hold, even in the frequency domain 
of the anomalous skin effect when, due to the spatial non-uniformity
of the field, the description in terms of $\varepsilon$ becomes impossible. 
Notice that in recent Ref.~\cite{46} it has been suggested to use
the ``exact'' boundary conditions (\ref{eq23}), (\ref{eq24}) with the
impedances depending on the wave vector (i.e. on the angle of incidence) 
instead of the Leontovich conditions
(\ref{eq26}) used in Refs.~\cite{35,36}. This, however, leads us back
to all the above problems with the thermal Casimir force, 
because in the absence of a mass-shell equation the representation for the
impedances (\ref{eq23}), (\ref{eq24}) becomes not equivalent to the
representation (\ref{eq25}). 
In our case we have used the Leontovich boundary condition (\ref{eq26});
that is to say, the representation (\ref{eq25}) was generalized for
the case of virtual photons. If one generalizes Eqs.~(\ref{eq23}),
(\ref{eq24}) for the case of virtual photons, this will lead to
zero value for the transverse reflection coefficient at zero frequency
and, therefore, to a contradiction with thermodynamics.

By the use of the surface impedance instead of the Drude model (\ref{eq3}),
the Lifshitz formula (\ref{eq1}) is preserved, but the coefficients
$r_{\|}^2$, $r_{\bot}^2$ given by Eq.~(\ref{eq2}) should be replaced 
by \cite{34}
\begin{equation}
r_{\|}^2(\zeta_l,y)=\left[\frac{y-Z(i\zeta_l)\zeta_l}{y
+Z(i\zeta_l)\zeta_l}\right]^2,
\quad
r_{\bot}^2(\zeta_l,y)=\left[\frac{\zeta_l-Z(i\zeta_l)y}{\zeta_l+
Z(i\zeta_l)y}\right]^2.
\label{eq27}
\end{equation}
\noindent
Substituting in Eq.~(\ref{eq27}) the impedance function of the normal
skin effect or the anomalous skin effect, one finds 
$r_{\|}^2(0,y)=r_{\bot}^2(0,y)=1$ \cite{35,36}. 
In the region of the infrared optics it follows
\cite{35,36} that 
$r_{\|}^2(0,y)=1$, 
$r_{\bot}^2(0,y)=({\tilde{\omega}}_p-y)^2/({\tilde{\omega}}_p+y)^2$.

It should be stressed that the expressions obtained for the reflection
coefficients at zero frequency in the impedance approach are exact.
They readily follow from the  exact Eq.~(\ref{eq25}) as
$|\sin^2\vartheta_0/\varepsilon|<(\gamma/\omega_p^2)\omega\to 0$
when $\omega\to 0$. This result is in contradiction with the
statement of Ref.~\cite{32a}. The authors of Ref.~\cite{32a}
start from representation (\ref{eq23}) for $Z_{\bot}$ and consider
the limit $\omega\to 0$ at fixed nonzero $k_{\bot}$ [i.e. they violate
the mass-shell equation from which Eqs.~(\ref{eq23}), (\ref{eq24})
are equivalent to Eq.~(\ref{eq25})]. As a result, the equality
$r_{\bot}^2(0,k_{\bot})=0$ obtained in Ref.~\cite{32a} leads
to the violation of the third law of thermodynamics (see Sec.~II).
In fact, both approaches, ours and that of Ref.~\cite{32a}, start
from different postulates. Our postulate is that the reflection
properties for the fluctuating field are the same as for real
photons. If this is true, the impedances (\ref{eq25}) follow,
which coincide with the Leontovich impedance at zero frequency
and are approximately equal to it at all nonzero frequencies with a very
high precision [all calculational results based on Eq.~(\ref{eq25})
and the Leontovich inpedance using $Z=1/\sqrt{\varepsilon}$ for metals
are practically the same]. In Ref.~\cite{32a} another postulate is
assumed, which admits that the reflection properties for the fluctuating
field are different from those for the usual electromagnetic field.
Both postulates have the right of being assumed because it is impossible
to study the reflection properties of the virtual photons experimentally.
The second postulate, however, is shown to be inconsistent with
thermodynamics. For this reason it must be rejected, and so we conclude
that the fluctuating field has the same reflection properties on
a metal boundary as the usual electromagnetic waves. 

Let us now present several computational results obtained using 
Eqs.~(\ref{eq1}) and (\ref{eq27}) 
in the framework of the impedance approach in comparison with the 
results calculated from the Lifshitz formula 
(\ref{eq1}), (\ref{eq2}) combined with the Drude model (\ref{eq3}). 
In Fig.~1, the magnitude of the Casimir force per unit area
$F=-\partial {\cal{F}}/\partial a$ for gold plates is plotted,
in the temperature interval 1\,K$\leq T\leq 1200\,$K at a separation
distance $a=1\,\mu$m. 
At this separation distance, the characteristic frequency $\omega_c$
belongs to the region of the infrared optics where the impedance
function is given by
\begin{equation}
Z(i\zeta_l)=\frac{\zeta_l}{\sqrt{{\tilde{\omega}}_p^2+\zeta_l^2}}.
\label{eq28a}
\end{equation}
\noindent
The solid line represents the values calculated in the framework of the
impedance approach, and the dashed line is obtained 
via the Lifshitz formula supplemented by the
Drude model with a temperature dependent relaxation parameter
[data for $\gamma(T)$ are taken from Ref.~\cite{47}].
It is clearly seen that the dashed line is not monotonous, demonstrating
the existence of a wide temperature region where the force modulus 
decreases with an increase of the temperature (as in Figs.~2,\,3 of
Ref.~\cite{32}). At the same time, the solid line obtained by using
the surface impedance demonstrates the monotonous
increase of the magnitude of the Casimir force with temperature in perfect
agreement with what is expected from thermodynamics.

It is instructive to know if the nonmonotonous force-temperature relation
takes place only for Drude metals or if this may also happen 
for dielectrics.
In Fig.~2, the magnitude of the Casimir force per unit area
 between dielectrics with $\varepsilon=\mbox{const}$ 
is shown for different temperatures at a separation
of $a=1\,\mu$m. Both solid and dotted
lines were obtained from the usual Lifshitz formula
(\ref{eq1}), (\ref{eq2}) with 
$\varepsilon(i\zeta_l)=\varepsilon=\mbox{const}$. 
The solid line is for mica
with $\varepsilon=7$; 
the dotted line coincides with the line in Fig.~5 of
Ref.~\cite{32} with $\varepsilon=100$.
The solid line shows a
monotonous increase of the Casimir force with the temperature, as is
expected from thermodynamics. On the other hand, the dotted line is not
monotonous as the dashed
line in Fig.~1. This is, however, artifactual because
$\varepsilon$ can be assumed to be independent on the frequency and 
temperature
only in the case of so-called non-polar dielectrics whose atoms
or molecules do not have their own dipole moments. The electric
susceptibility of non-polar dielectrics arises due to the electronic
polarization of atoms and molecules. The values of 
$\varepsilon$ for non-polar
dielectrics are of the order of one \cite{47,48,49}.
Large values of $\varepsilon$ can exist only for polar dielectrics where
the partial orientation of permanent dipole moments occurs.
But for polar dielectrics 
$\varepsilon$ depends strongly on the frequency and 
temperature. Specifically, their 
$\varepsilon$ quickly decreases with the increase 
of frequency. As a result, at optical and infrared frequencies, which
are characteristic for the Casimir effect, the values of $\varepsilon$
are determined by the electronic polarizability \cite{47} and cannot
exceed several units.
The dielectric permittivity of the polar dielectrics along the imaginary
frequency axis can be modeled by \cite{50}
\begin{equation}
\varepsilon(i\xi)=1+\sum\limits_{n=1}^{N}
\frac{C_n}{1+\frac{\xi}{\omega_n}},
\label{eq29}
\end{equation}
\noindent
where the sum describes the effect of possible Debye rotational
relaxation frequencies (the absorption spectra
of dielectrics are not influential at the characteristic frequencies
of the Casimir effect $\omega_c$ when the separation between plates
is of the order of $1\,\mu$m). In solids, the polarization due to
orientation of the permanent dipole moments disappears at rather
low frequencies $\sim(10^{7}-10^{8})\,$rad/s \cite{44}.
At very high frequencies $\omega>10^{16}\,$rad/s, $\varepsilon$
decreases to unity \cite{40}. Therefore, for the model calculation 
we may choose $N=2$, $C_1=93$, $\omega_1=10^{7}$\,rad/s,
$C_2=6$, $\omega_2=10^{16}$\,rad/s.  In this case 
$\varepsilon(0)=100$ and $\varepsilon\approx 7$ at the characteristic
frequency $\omega_c$ corresponding to $a=1\,\mu$m.

Eq.~(\ref{eq29}) was substituted into the Lifshitz formula
(\ref{eq1}), (\ref{eq2}). In Fig.~2,
the magnitude of the Casimir force per unit area between the polar
dielectrics is represented by the dashed line. It is seen that the
force-temperature relation, given by the dashed line, is monotonous
as in the case of non-polar dielectrics with small $\varepsilon$,
as expected from thermodynamics.
{}From Fig.~2 it is clear that the 
Casimir force between dielectrics is 
a monotonous function of the temperature 
when realistic input data for $\varepsilon$ are substituted into
the Lifshitz formula. 

In Fig.~3, the Casimir entropy for gold is plotted as a function
of the temperature at a separation distance between plates
$a=1\,\mu$m. The solid line is drawn in accordance with the
impedance approach, i.e. by the use of Eqs.~(\ref{eq1}), (\ref{eq27}) 
and (\ref{eq28a}). The dashed line is obtained from the usual Lifshitz 
formula and Drude model (\ref{eq1})--(\ref{eq3}).
Evidently, the solid line satisfies all conditions, 
i.e. positive values of the entropy at nonzero temperatures, and
the validity of the Nernst heat theorem. By contrast, the dashed
line presents the negative values of the entropy and the violation
of the Nernst heat theorem. 
The analytical proof of the validity of the Nernst heat theorem 
in the impedance approach can be found in Ref.~\cite{36}.

\section{Conclusions and discussion}

As we have proved above, the substitution of the Drude dielectric function
into the Lifshitz formula for the thermal Casimir force leads to the
violation of the third law of thermodynamics (the Nernst heat theorem).
A rigorous analytical evidence of this statement lies on the fact
that at low temperatures the magnitude of the relaxation
parameter is much less than the Matsubara frequencies, a property
which is always true in the case of perfect crystal lattice.
A special analysis of the role of defects or impurities leads to
the conclusion that they are incapable to reconcile the calculations
of the thermal Casimir force between Drude metals with
thermodynamics.

It has been known that the Lifshitz formula combined with the
Drude dielectric function predicts a linear (in temperature) large
thermal correction to the Casimir force at short separations
connected with the second term in the free energy (\ref{eq8})
\cite{27,32,33,38}. In recent precision measurement
of the Casimir pressure between Au and Cu plates at room
temperature \cite{15,16} this correction (which we call
``alternative'' \cite{16}) comprises 4.89\,mPa and 1.23\,mPa
at separations $a=300\,$nm and 500\,nm, respectively, i.e.
3.6\% and 7.24\%, respectively, of the total Casimir pressure.
For reference, the traditional thermal corrections in the same 
experimental configuration (computed using the plasma dielectric
function \cite{25,26} or the surface impedance \cite{35,36}) at
separations $a=300\,$nm and 500\,nm are equal to
--0.00863\,mPa and --0.00441\,mPa, respectively, i.e. only
0.006\% and 0.03\%, respectively, of the total Casimir pressure
(note that the traditional thermal corrections are of the same
sign as the Casimir force, i.e. negative).
The experimental results of Refs.~\cite{15,16} and the extent
of their agreement with theory rule out the linear thermal
correction to the Casimir force as predicted by the Drude
dielectric function substituted into the Lifshitz formula
(see Ref.~\cite{16} for details). Thus, the combination of the
Drude model with the Lifshitz formula is not only thermodynamically
inconsistent, but is also in contradiction with experiment.

As we have demonstrated through several computations, the description
of real metals based on the surface impedance (Leontovich
boundary conditions) is thermodynamically consistent.
Unlike the dielectric permittivity, 
the surface impedance is well defined at all frequencies, even in the
domain of the anomalous skin effect. By the use of the
Leontovich impedance boundary conditions, 
instead of the Drude model, the Lifshitz formula is preserved,
but the reflection coefficients are expressed in terms of the
impedance function. 
An important point is that different representations for the 
impedances, which are equivalent for real photons, become nonequivalent
in application to the fluctuation fields. As we have shown above, in the
case of fluctuating fields the representation (\ref{eq25}) for the
impedance should be used leading to the Leontovich boundary
conditions, whereas the representations (\ref{eq23}), (\ref{eq24}),
when applied to virtual photons, lead to the contradictions
with thermodynamics.

In the framework of the impedance approach the value of the 
zero-frequency term of the Lifshitz formula is prescribed by the 
form of the impedance and quite satisfactory
physical results are obtained. In particular, the Casimir energy and
force turn out to be monotonous functions of the temperature, 
in agreement with what would be expected from thermodynamics. 
In the region of the infrared optics the results obtained by 
using the surface impedance coincide with those found earlier 
through the use of the plasma dielectric function \cite{25,26}.
The Casimir
entropy in the impedance approach is always positive and vanishes
at zero temperature, in accordance with the Nernst heat theorem.

To conclude, we have proved that
the Drude dielectric function is not appropriate to describe
the thermal Casimir effect in the case of real metals, leading
to contradictions with thermodynamics. 
On the other hand, the
Lifshitz formula with the coefficients expressed in terms of the surface
impedance is suitable to calculate all the quantities
of physical interest.

\section*{Acknowledgments}

The authors are grateful to CNPq and Finep for partial
financial support.

\begin{figure*}
\vspace*{-6cm}
\includegraphics{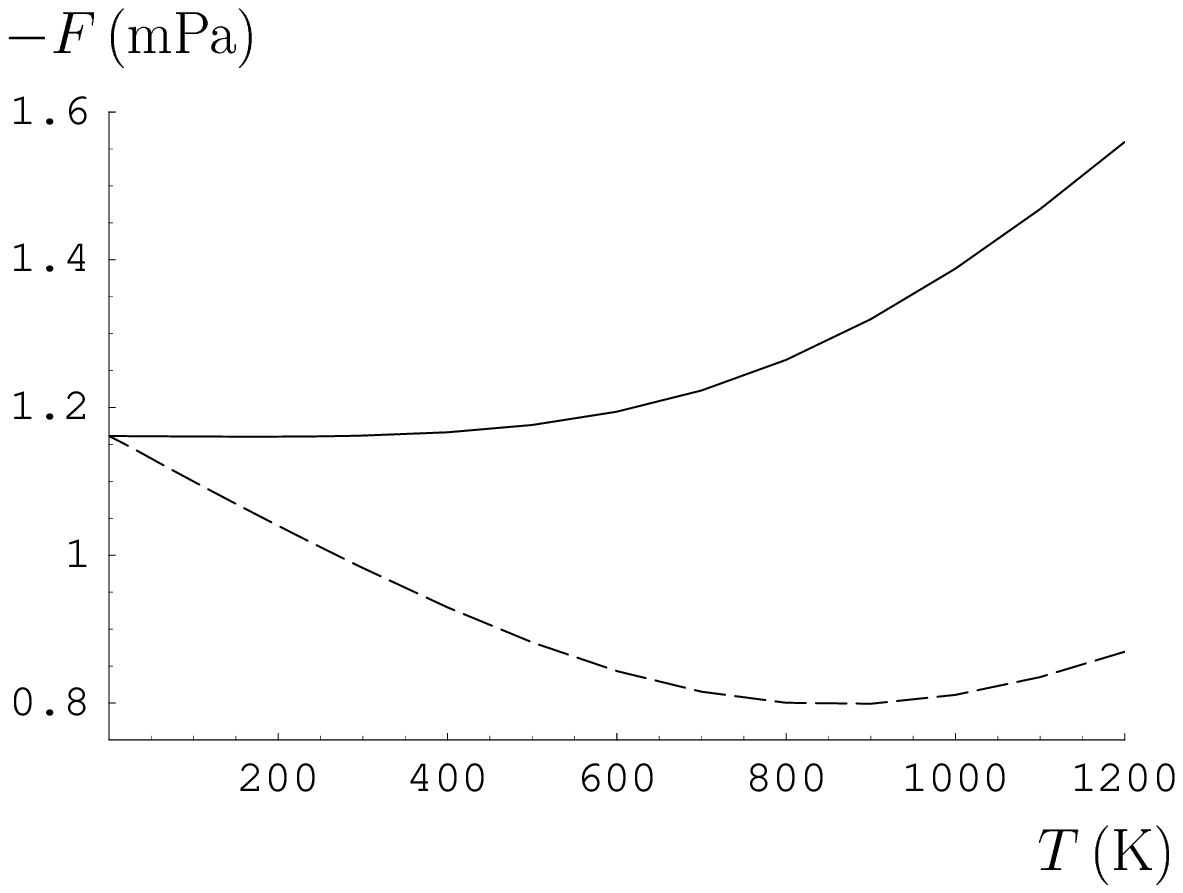}
\vspace*{-7cm}
\caption{Magnitude of the Casimir force per unit area for gold 
versus temperature when $a=1\,\mu$m. The solid line is obtained
in the framework of the impedance approach.
The dashed line is obtained by using the Drude dielectric
function.
}
\end{figure*}
\begin{figure*}
\vspace*{-7cm}
\includegraphics{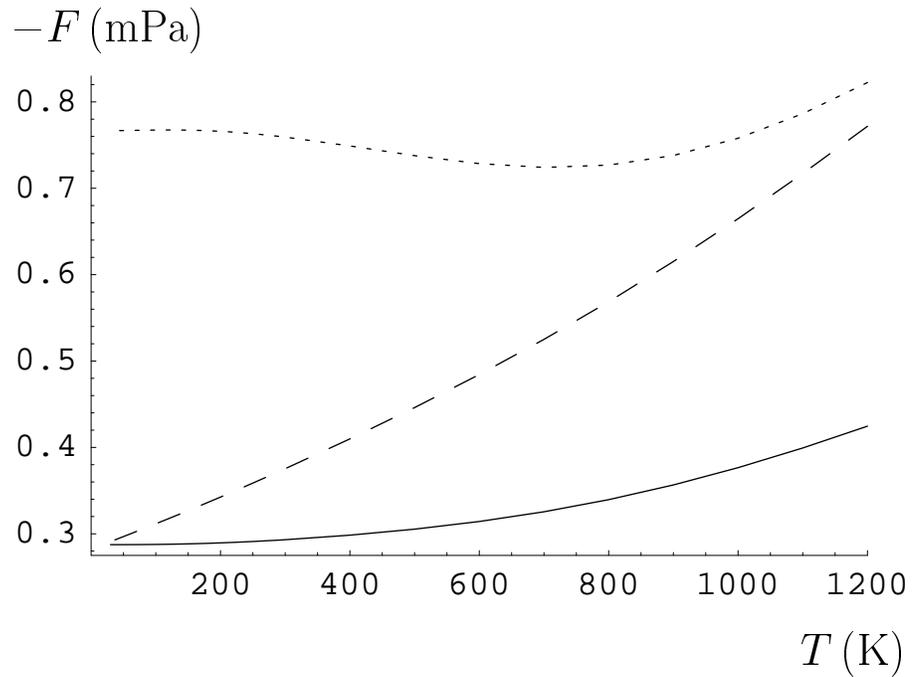}
\vspace*{-8cm}
\caption{Magnitude
of the Casimir force per unit area for 
dielectrics versus
temperature when $a=1\,\mu$m, calculated by the Lifshitz formula. 
The solid line is  for mica, with $\varepsilon=7$.
The dotted line is obtained for non-existent non-polar
dielectric with $\varepsilon =100$.
The dashed line is for polar dielectric with $\varepsilon(0)=100$.
}
\end{figure*}
\begin{figure*}
\vspace*{-7cm}
\includegraphics{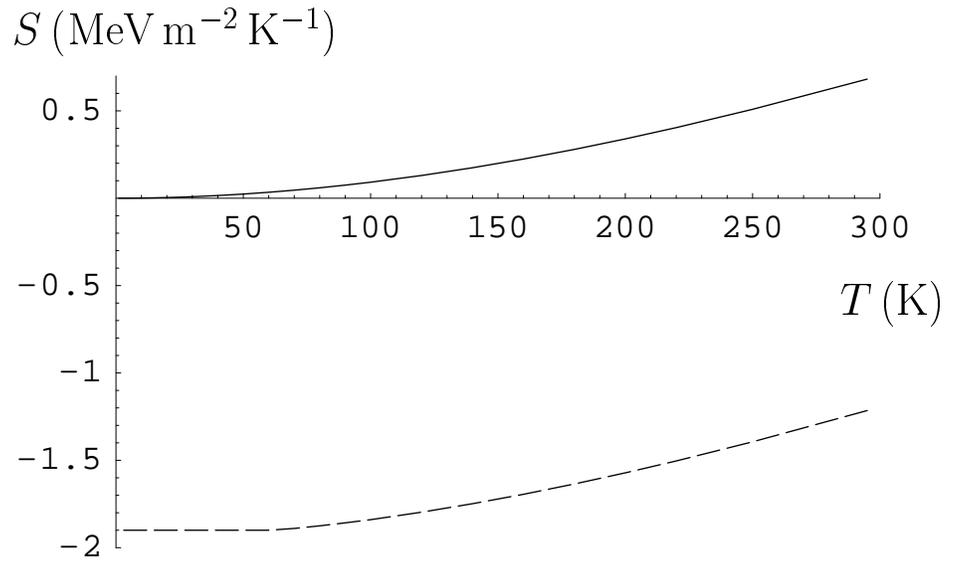}
\vspace*{-8cm}
\caption{Casimir entropy per unit area
for two gold semispaces versus temperature,
when $a=1\,\mu$m. The solid line is obtained
in the framework of the impedance approach.
The dashed line is obtained by using the Drude dielectric
function.
}
\end{figure*}
\end{document}